\documentclass[12pt]{article}
\usepackage{mathrsfs}
\usepackage{amssymb}
\usepackage{amsmath}
\usepackage[amsmath,thmmarks]{ntheorem}
\usepackage{fullpage}

\usepackage{graphicx}
\usepackage{bm}
\usepackage{slashed}
\usepackage{color}

\def\lsim{\mathrel{\rlap{\lower4pt\hbox{\hskip1pt$\sim$}}
    \raise1pt\hbox{$<$}}}                
\def\gsim{\mathrel{\rlap{\lower4pt\hbox{\hskip1pt$\sim$}}
    \raise1pt\hbox{$>$}}}                

\def\OMIT#1{}

\newcommand{\be}{\begin{eqnarray}}
\newcommand{\ee}{\end{eqnarray}}

\newcommand{\nn}{\nonumber}
\newcommand{\w}{\omega}

\newcommand{\bn}{{\bar n}}
\newcommand{\bea}{\begin{eqnarray}}
\newcommand{\eea}{\end{eqnarray}}

\newcommand{\bnp}{\bar n \!\cdot\! p}
\newcommand{\bnP}{\bar {\cal P}}

\def\lsim{\mathrel{\rlap{\lower4pt\hbox{\hskip1pt$\sim$}}
    \raise1pt\hbox{$<$}}}                
\def\gsim{\mathrel{\rlap{\lower4pt\hbox{\hskip1pt$\sim$}}
    \raise1pt\hbox{$>$}}}                

\def\OMIT#1{}

\begin{document}

\begin{titlepage}

\makebox[6.5in][r]{\hfill ANL-HEP-PR-12-76}

\vskip1.40cm
\begin{center}
{\Large {\bf Resummation of jet-veto logarithms in hadronic processes containing jets}} \vskip.5cm
\end{center}
\vskip0.2cm

\begin{center}
{\bf Xiaohui Liu, Frank Petriello}
\end{center}
\vskip 6pt
\begin{center}
{\it Department of Physics \& Astronomy, Northwestern University, Evanston, IL 60208, USA} \\
{\it High Energy Physics Division, Argonne National Laboratory, Argonne, IL 60439, USA} \\
\end{center}

\vglue 0.3truecm

\begin{abstract}
We derive a factorization theorem for production of an arbitrary number of color-singlet particles accompanied by a fixed number of jets at the LHC.  The jets are defined with the standard anti-$k_T$ algorithm, and the fixed number of jets is obtained by imposing a veto on additional radiation in the final state. The formalism presented here is useful for current Higgs boson analyses using exclusive jet bins, and for other studies using a similar strategy.  The derivation uses the soft-collinear effective theory and assumes that the transverse momenta of the hard jets are larger than the veto scale.  We resum the large Sudakov logarithms $\alpha_s^n \log^{2n-m}\left(p_T^{J}/p_T^{veto}\right)$ up to the next-to-leading-logarithmic accuracy, and present numerical results for Higgs boson production in association with a jet at the LHC.  We comment on the experimentally-interesting parameter region in which we expect our factorization formula to hold.

\end{abstract}

\end{titlepage}
\newpage

\section{Introduction}

Accurate predictions for processes with a fixed number of final-state jets are crucial for many LHC applications.  A well-known example is that of a Higgs boson decaying to $W$-bosons at the LHC~\cite{Aad:2012npa,Chatrchyan:2012ty}. The background composition to this signal changes as a function of jet multiplicity.  In the zero-jet bin the background is dominated by continuum $WW$ production, while in the one-jet and two-jet bins, top-pair production becomes increasingly important.  The optimization of this search requires cuts dependent on the number of jets observed, and therefore also on theoretical predictions for exclusive jet multiplicities.  

Theoretical predictions for processes with an exclusive number of jets are notoriously difficult to obtain.  Fixed-order perturbation theory is plagued by large logarithms of the form $\text{ln} (Q/p_T^{veto})$, where $Q$ denotes the hard scale 
in the process, such as $m_H$.  For experimentally relevant values $p_T^{veto} \sim 25-30$ GeV, residual scale variations in fixed-order calculations lead to estimated errors that do not accurately reflect uncalculated higher-order corrections~\cite{Anastasiou:2008ik,Stewart:2011cf,Banfi:2012yh}.  Progress in resummation of these logarithms to all orders has been slow.  Event-shape variables such as jettiness~\cite{Stewart:2010tn} allow resummation of jet-veto effects to arbitrary logarithmic accuracy, and have been applied to study the production of vector bosons or Higgs bosons plus multiple jets at the LHC~\cite{Liu:2012zg,Berger:2010xi}.  However, experimental measurements typically utilize jet algorithms such as the anti-$k_T$ algorithm, and conclusions drawn from calculations using jettiness necessarily remain qualitative only.  Resummation of jet-veto logarithms for the Higgs cross section in the zero-jet bin in the presence of the anti-$k_T$ algorithm has been performed at next-to-leading logarithmic (NLL) accuracy using the semi-numerical program {\tt CAESER}~\cite{Banfi:2012yh}\footnote{We note that different schemes for counting logarithms are employed in the literature; we specify in detail the order-counting scheme we use in Section~\ref{secfact}.}.  Recent work has extended these results to their NNLL accuracy~\cite{Becher:2012qa,Banfi:2012jm}.  It has been pointed out the potentially large $ \text{ln} \,R$ corrections, where $R$ is the jet-radius parameter in the anti-$k_T$ algorithm, could have a significant numerical impact on the predictions~\cite{Tackmann:2012bt}.  These terms have yet to be studied at all orders and warrant further investigation.

We consider in this manuscript the resummation of the jet-veto logarithms for production of one or more color-neutral particles, such as a Higgs boson or electroweak gauge bosons, in association with one or more jets.   We accomplish this by deriving a factorization theorem using soft-collinear effective theory (SCET)~\cite{Bauer:2000ew,Bauer:2000yr,Bauer:2001ct,Bauer:2001yt,Bauer:2002nz} that assumes that the transverse momenta of the hard jets are larger than the veto scale.  As an example application, we consider explicitly Higgs boson production in association with a single jet.  This calculation is of direct phenomenological interest for understanding the properties of the new Higgs-like state observed at the LHC~\cite{:2012gk,:2012gu}.  It extends previous work on understanding the effect of resummation on the Higgs plus zero-jet cross section~\cite{Becher:2012qa,Banfi:2012jm,Tackmann:2012bt}.  We resum the logarithms $\text{ln} (Q/p_T^{veto})$ through the next-to-leading logarithmic (NLL) level, where $Q \sim m_H \sim p_T^J$ and $p_T^J$ is the transverse momentum of the observed jet.  We demonstrate that the residual scale variation of the theoretical prediction is drastically reduced by the inclusion of the NLL resummation, and that the NLL+NLO result provides reliable predictions over a larger kinematic range.  Since the factorization theorem we derive is valid for both more jets and other color-neutral particles, our result also serves as a framework for how to augment a host of fixed-order calculations with resummation of a class of large logarithms.  

In the context of recent work suggesting that it is difficult to extend resummation of jet-veto logarithms in the presence of the anti-$k_T$ algorithm beyond the NLL level~\cite{Tackmann:2012bt}, a relevant question to consider is the required accuracy of the resummation to match experimental needs.  Phrased more colloquially, how many N's of logarithmic accuracy are needed?  In the example above, the numerical value of the 
leading logarithmic term in the experimentally interesting region is $\text{ln}^2 (Q/p_T^{veto}) \approx 2.5$.  This is not an overwhelmingly large correction.  The ability to supplement fixed-order codes with simple analytical resummation of important sources of logarithms to the NLL level should be sufficient.  This would allow these programs to be extended near regions of phase space where large logarithms appear, while still producing reliable central values and theoretical error estimates within a controlled theoretical approximation.  We comment in more detail later in this manuscript on the numerical relevance of missing higher-order corrections that may arise at the NNLL level.  We find that in the experimentally interesting parameter region for Higgs production, it is justified to resum the large logarithms associated with the jet veto.  
  
Our paper is organized as follows. In Section~\ref{secfact}, we 
derive our factorization theorem using SCET. We apply our
formalism to study Higgs plus one-jet production, and discuss our numerical results, in Section~\ref{sechiggs}.  We conclude and discuss future directions in Section~\ref{seccon}. All formulae needed for resummation at the NLL level are given in the Appendix.

\section{Factorization and Resummation}\label{secfact}

In this section we derive a factorization theorem for multi-jet production at the LHC in the presence of a jet veto.  We discuss the resummation of the logarithms associated with the jet veto through NLL accuracy.  We use $pp \to {\rm Higgs} + 1\ {\rm jet}$ via gluon-gluon fusion as an example to highlight the derivation procedure.  The generalization to additional jets is straightforward, and is presented here.  Our primary results are contained in Eqs.~(\ref{facthiggs}),~(\ref{factgen}),~(\ref{resumhiggs}) and~(\ref{resumgen}).

\subsection{Discussion of the jet constraints}

We focus on the case in which the jets are defined using the hadron collider
anti-$k_T$ algorithm~\cite{Cacciari:2008gp}.  The following distance metrics are used:
\begin{eqnarray}
\rho_{ij} &=& {\rm min}(p_{T,i}^{-1},p_{T,j}^{-1})\Delta R_{ij}/R, \nonumber \\
\rho_i &=& p_{T,i}^{-1}. 
\end{eqnarray}
The anti-$k_T$ algorithm merges particles $i$ and $j$ to form a new particle by adding their four momenta
if $\rho_{ij}$ is the smallest among all the metrics.  Otherwise, $i$ or $j$ is promoted to a jet depending on whether $\rho_i$ or $\rho_j$ is smaller, and removed from the set of considered particles.  This procedure is repeated until all particles are grouped into jets.  We note that $\Delta R_{ij}^2 = \Delta \eta_{ij}^2 + \Delta \phi_{ij}^2$,
where $\Delta \eta_{ij}$
and $\Delta \phi_{ij}$ are the rapidity and azimuthal angle
difference between particles $i$ and $j$, respectively. $R$ is the
jet radius parameter and in practice is chosen to be around $0.4-0.5$. 

After clustering the partons, we demand that the final state contains no additional jets with transverse momentum greater than a threshold $p_T^{veto}$.  For Higgs searches, typically $p_T^{veto} \sim 25-30$ GeV.  Since $p_T^{veto}$ is usually substantially lower than the partonic center of mass ($\lambda \equiv p_T^{veto}/\sqrt{\hat{s}} \ll 1$), 
the vetoed observables are usually very sensitive to soft and collinear emissions.

Additional constraints beyond the jet veto can be imposed on the final state. In the following derivation, we require that the measured leading jets are all well-separated so that no additional
small scales will be generated. We also assume that leading jet momentum $p_T^J \sim m_H \sim \sqrt{\hat{s}}$ and 
$1\gg R^2\gg \lambda^2 $ while $\frac{\alpha_s}{2\pi} \log^2 R \ll 1$.  The second of these requirements is necessary to insure that the measurement function factorizes into separate measurements in each of the collinear sectors.  This is discussed in detail in the next section.  The third requirement ensures that logarithms associated with the anti-$k_T$ parameter $R$ need not be resummed.  Given that $p_T^{veto} \approx 25-30{\rm GeV}$ and $R \approx 0.4-0.5$ 
for Higgs production,when the leading jet $p_T^J \approx m_H \approx 125{\rm GeV}$, these assumptions are justified. 

\subsection{Derivation of the factorization theorem}

We use SCET~\cite{Bauer:2000ew,Bauer:2000yr,Bauer:2001ct,Bauer:2001yt,Bauer:2002nz} to establish our factorization theorem.  SCET makes manifest the infrared limits of QCD by re-formulating 
the QCD Lagrangian using soft and collinear modes whose momenta scale with a small power-counting parameter $\lambda$ in appropriate ways.  For Higgs production, this parameter is of order $p_T^{veto}/\sqrt{\hat{s}}$ for radiation outside
the measured jet, and is of order $R$ for radiation inside the measured jet.  Consideration of the jet algorithm and jet veto lead to the following relevant degrees of freedom in the effective theory:
\begin{itemize}

\item a collinear jet mode with momentum 
$p_J = \frac{\w_J}{2} n_J + k_J$, where $n_J$ is the light-cone vector along the jet direction;

\item two collinear modes propagating along the beam
axes $a$ and $b$, with $p_i = \frac{\w_i}{2}n_i + k_i$ for $i = a,b$;

\item a soft mode with momentum $k_s$.

\end{itemize}
The residual momenta $k_J$, $k_i$
and the soft momentum $k_s$ all scale as $\sqrt{\hat{s}} \lambda$, while
 the large components of the three collinear momenta scale as $\w_i \sim \sqrt{\hat{s}}$.  We note that no ultrasoft fields ($k_{us}\sim \sqrt{\hat{s}} \lambda^2$) are needed for the process
we are considering here.  Any phase space measurement $\hat{{\cal M}}$
is assumed to be insensitive to these modes in the final state, so that 
$\sum_{us}\hat{{\cal M}}|X_{us}\rangle \langle X_{us}| = 
\sum_{us}|X_{us}\rangle \langle X_{us}| = 1 +{\cal O}(\lambda)
$, where $X_{us}$ denotes the final-state particles with an ultrasoft momentum scaling.

The leading-power SCET operator which mediates gluon-fusion Higgs plus one jet production is 
\bea
H{\cal O}(x) = \sum_{\w_i,n_i}
e^{-i\left(\frac{\w_1}{2}n_1 + \frac{\w_2}{2}n_2 - \frac{\w_3}{2}n_3 \right)\cdot x} \,
C_{\alpha\beta\mu}^{abc} H \,
S^{aa'}_{n_1}S^{bb'}_{n_2}S^{cc'}_{n_3}
{\cal B}_{\w_1,n_1}^{\alpha,a'}{\cal B}_{\w_2,n_3}^{\beta,b'}{\cal B}_{\w_3,n_3}^{\mu,c'}(x)\,,
\eea
where we have explicitly written out the Lorentz indices $\alpha\beta\mu$ and
the color indices $abc$.  $H$ is the operator which creates
a Higgs boson in the final state. $C_{\alpha\beta\mu}^{abc}$ is the hard Wilson coefficient which encodes
the hard virtual fluctuations.  It can be obtained order-by-order by calculating
the corresponding QCD diagrams. The $n$-collinear gauge invariant boson field~\cite{Bauer:2002nz}
\bea
{\cal B}^\alpha_{n,\w} = \frac{1}{g_s}\left[W^\dagger_n({\cal P}^\alpha_\perp + g_s A^\alpha_{n\perp})W_n\right] \delta_{\w,{\cal \bar{P}}}\,
\delta_{n,\hat{n}}\,,
\eea
which creates (annihilates) a collinear gluon in the final (initial) state, is built out of the collinear Wilson line~\cite{Bauer:2001ct}
\bea
W_n=\sum_{\rm perm.} \exp\left[-g_s\frac{1}{\cal {\bar P}}\bn\cdot A_n \right]\,.
\eea
${\cal P}$ is the projective operator acting on the collinear fields
sitting to the right of it inside the parentheses. At leading power, the
interactions between collinear and soft fields can be eliminated through an
operator redefinition ${\cal B}^a_n \to S^{aa'}_n {\cal B}^{a'}_n $, which results in the soft Wilson line $S_n^{aa'}$~\cite{Bauer:2001yt}.

The cross section with a jet veto can be written as
\bea
\frac{\mathrm{d}\sigma}{ \mathrm{d} \Phi_H } =\frac{1}{8s}\sum_{spin} \sum_X
\int \mathrm{d}x \, e^{-iq_H\cdot x}
\langle p_ap_b| {\cal O}^\dagger(x){\cal \hat{M}}|X_aX_bX_JX_s\rangle \langle X_sX_JX_bX_a|{\cal O}(0) |p_ap_b\rangle \,.
\eea
We have decomposed the final state into different sectors based on
the momentum scaling. The operator ${\cal O}$ has been written
in a factorizable form, but in order to proceed, we must demonstrate that the measurement operator 
${\cal \hat{M}}$, which includes the jet clustering operation and jet vetoing,
can also be factored into different sectors up to power suppressed corrections~\cite{Walsh:2011fz}.  The factorizability of the measurement operator $\hat{{\cal M}}$ can be seen
through power-counting the anti-$k_T$ algorithm metrics $\rho_{ij}$ and $\rho_i$. Recalling that the transverse momentum $p_T$ for each sector
scales as
\bea
p_T^J \sim {\cal O}(1)\,,
\hspace{3.ex}
p_T^s \sim p_T^a \sim p_T^b \sim {\cal O}(\lambda)\,,
\eea
we have
\bea
&&\rho_{JJ} \lesssim \rho_J\sim  1\,, 
\hspace{3.ex}
\rho_{Js} \sim R^{-1}\,,
\hspace{3.ex}
\rho_{Ja} \sim \rho_{Jb} \sim R^{-1}\log \lambda^{-1}\,, \nn\\
&&\rho_{ss} \sim \rho_{aa}\sim \,
\rho_{bb} \sim (\lambda R)^{-1}\,,
\hspace{3.ex}
\rho_{sa}\sim \rho_{sb} \sim \rho_{ab} \sim  (\lambda R)^{-1} \log\lambda^{-1} \,, \nn\\
&& \rho_s \sim \rho_a \sim \rho_b \sim \lambda^{-1} \,.
\eea
These scalings indicate that for the jet-radius parameter $R$ not too large
($R \ll \log \lambda^{-1} $), the contributions from
the mixing between the jet and beam sectors, and the mixing between the soft and beam sectors, are power suppressed.  Jets tend to form separately within each sector.  As long as $R \ll 1$, radiation collinear to the jet direction will be combined before it is clustered with soft radiation.  This means that the soft radiation is insensitive
to the details of the collinear radiation except for the jet direction.  Therefore, $\hat{\cal M}$ can also be factored between these two sectors.  The measurement operator can therefore be factored as 
\bea
\hat{\cal M} = \hat{\cal M}_J\hat{\cal M}_s\hat{\cal M}_a\hat{\cal M}_b,
\eea
up to power-suppressed corrections in $p_T^{veto}$ and $R$.  The individual ${\cal M}_A$ will be given in the Appendix.

Plugging in the definition of the operator ${\cal O}$, the cross section can be written as
\bea
\frac{\mathrm{d}\sigma}{\mathrm{d} \Phi_H}
& =& \frac{1}{8s}  \sum_{spin} \sum_{n_A}\int\mathrm{d} \w_A 
\int \mathrm{d}x \, e^{i\left(\frac{\w_a}{2}n_a + \frac{\w_b}{2}n_b - \frac{\w_J}{2}n_J -q_H\right)\cdot x}\nn\\
&&\hspace{5.ex}\times C_{\alpha\beta\mu}^{abc\dagger}C_{\alpha'\beta'\mu'}^{a'b'c'}
\sum_{X_s}'\langle 0| S^{aa_1}_{n_a}S^{bb_1}_{n_b}S^{cc_1}_{n_J}(x)|X_s \rangle\,
 \langle X_s | S^{a_1'a'}_{n_a}S^{b_1'b'}_{n_b}S^{c_1'c'}_{n_J}(0)|0\rangle \nn \\
&&\hspace{5.ex}\times \,
\sum_{X_a}'
\langle p_a|{\cal B}_{n_a}^{\alpha,a_1}(x^-_{n_a},x^\perp_{n_a},0_{n_a})|X_a\rangle
 \langle X_a|\delta(\w_a-\bnP_a)\delta_{n_a,\hat{n}_a}{\cal B}_{n_a}^{\alpha',a_1'}(0) |p_a\rangle \nn\\
&&\hspace{5.ex}\times\,
\sum_{X_b}'
\langle p_b|{\cal B}_{n_b}^{\beta,b_1}(x^-_{n_b},x^\perp_{n_b},0_{n_b})|X_b\rangle
 \langle X_b|\delta(\w_b-\bnP_b)\delta_{n_b,\hat{n}_b}{\cal B}_{n_b}^{\beta',b_1'}(0) |p_b\rangle
\nn \\
&&\hspace{5.ex} \times\,
\sum_{n_J}\sum_{X_J}'
\langle 0|{\cal B}_{n_J}^{\mu,c_1}(x^-_{n_J},x^\perp_{n_J},0_{n_J}) |X_J\rangle 
\langle X_J|\delta(\w_J-\bnP_J)\delta_{n_J,\hat{n}_J}{\cal B}_{n_J}^{\mu',c_1'}(0) |0 \rangle \,.
\eea
Here, $\sum_X'$ means summing over the final states with the restrictions imposed by $\hat{\cal M}$.  We have suppressed $\hat{\cal M}$ for simplicity. In the last line, $\sum_{n_J}$ indicates the need to sum over all possible directions of different $n_J$-jet modes.
We have removed several Kronecker-deltas using the discrete sums.  The remaining ones
have been turned into integrals $\int \mathrm{d} \w_A$ and delta functions $\delta(\w_A-\bar{\cal P}_A)$, after combining the residual momentum $k^-_{n_A} \equiv \bn_A \cdot k_A$ with the label momentum $\w_A$
in the collinear sector $A$ for $A =a,b,J$.  We note that the collinear sectors do not depend on $x_A^+ \equiv n_A\cdot x$.

We further modify this expression by performing a translational operation in each
sector and inserting several residual momentum operators $\int \mathrm{d}k \delta(k-\hat{k})$ to remove the explicit momentum dependence on the final states. This gives us
\bea
\frac{\mathrm{d}\sigma }{ \mathrm{d} \Phi_H }
& =& \frac{1}{8s}\sum_{spin} \sum_{n_A}\int\mathrm{d} \w_A 
\int \mathrm{d}x \int\mathrm{d}^4k_s \int\mathrm{d}{\bf k}_A \,
 e^{i\left(\frac{\w_a}{2}n_a + \frac{\w_b}{2}n_b - \frac{\w_J}{2}n_J \,
-q_H-k_s + {\bf k}_a + {\bf k}_b -{\bf k}_j\right)\cdot x}\nn\\
&&\hspace{5.ex}\times C_{\alpha\beta\mu}^{abc\dagger}C_{\alpha'\beta'\mu'}^{a'b'c'}\,
\sum_{X_s}'
\langle 0| S^{aa_1}_{n_a}S^{bb_1}_{n_b}S^{cc_1}_{n_J}(0)|X_s \rangle\,
 \langle X_s |\delta^4(k_s-\hat{k}_s) S^{a_1'a'}_{n_a}S^{b_1'b'}_{n_b}S^{c_1'c'}_{n_J}|0\rangle \nn \\
&&\hspace{5.ex}\times \,
\sum_{X_a}'\,
\langle p_a|{\cal B}_{n_a}^{\alpha,a_1}(0)|X_a\rangle
 \langle X_a|\delta(k_a^+-\hat{k}_a^+)\delta^2(k_a^\perp - \hat{k}_a^\perp)\delta(\w_a-\bnP_a)\delta_{n_a,\hat{n}_a}{\cal B}_{n_a}^{\alpha',a_1'}(0) |p_a\rangle \nn\\
&&\hspace{5.ex}\times\,
\sum_{X_b}'\,
\langle p_b|{\cal B}_{n_b}^{\beta,b_1}(0)|X_b\rangle
 \langle X_b|\delta(k_b^+-\hat{k}_b^+)\delta^2(k_b^\perp - \hat{k}_b^\perp)\delta(\w_b-\bnP_b)\delta_{n_b,\hat{n}_b}{\cal B}_{n_b}^{\beta',b_1'}(0) |p_b\rangle
\nn \\
&&\hspace{5.ex} \times\,
\sum_{X_J}'\,
\langle 0|{\cal B}_{n_J}^{\mu,c_1}(0) |X_J\rangle 
\langle X_J|\delta(k_J^+-\hat{k}_J^+)\delta^2(k_J^\perp - \hat{k}_J^\perp)\delta(\w_J-\bnP_J){\cal B}_{n_J}^{\mu',c_1'}(0) |0 \rangle \,,
\eea
where $\mathrm{d}{\bf k} \equiv \mathrm{d}k^+\mathrm{d}^2k^\perp$. 
We now drop all residual momenta of order $\lambda$ or higher, keeping homogeneously only the leading power terms in the exponent (dropping the contributions of order $\lambda$ means that we ignore the recoil effect in the transverse plane). 
We perform the integration over the $k^+_A$ component in each 
collinear sector to indicate that no restrictions are applied 
on this residual momentum~\footnote{We note that this assumes that no pseudorapidity cut is imposed on the observed jet.  It is straightforward to remove this constraint if desired; for simplicity of presentation we do not do so here.}. 
This leads to  
\bea
\frac{\mathrm{d}\sigma }{\mathrm{d} \Phi_H }
& =& \frac{1}{8s} \sum_{spin} \int\mathrm{d} \w_A\,
\sum_{n_J} \mathrm{d}^2k_J^\perp  \,
 (2\pi)^4 \delta^4\left(\frac{\w_a}{2}n_a + \frac{\w_b}{2}n_b - \frac{\w_J}{2}n_J \,
-q_H\right)\nn\\
&&\hspace{5.ex}\times C_{\alpha\beta\mu}^{abc\dagger}C_{\alpha'\beta'\mu'}^{a'b'c'}\,
\int\mathrm{d}^4k_s\langle 0| S^{aa_1}_{n_a}S^{bb_1}_{n_b}S^{cc_1}_{n_J}(0)
\hat{\cal M}_s
\delta^4(k_s-\hat{k}_s) S^{a_1'a'}_{n_a}S^{b_1'b'}_{n_b}S^{c_1'c'}_{n_J}|0\rangle \nn \\
&&\hspace{5.ex}\times
 \int\mathrm{d}^2 k^\perp_a
 \langle p_a|{\cal B}_{n_a}^{\alpha,a_1}(0)\,
\hat{\cal M}_a\,
\delta^2(k_a^\perp - \hat{k}_a^\perp)\delta(\w_a-\bnP_a){\cal B}_{n_a}^{\alpha',a_1'}(0) |p_a\rangle \nn\\
&&\hspace{5.ex}\times
 \int\mathrm{d}^2k^\perp_b
\langle p_b|{\cal B}_{n_b}^{\beta,b_1}(0)\,
\hat{\cal M}_b\,
\delta^2(k_b^\perp - \hat{k}_b^\perp)\delta(\w_b-\bnP_b){\cal B}_{n_b}^{\beta',b_1'}(0) |p_b\rangle
\nn \\
&&\hspace{5.ex} \times\,
\langle 0|{\cal B}_{n_J}^{\mu,c_1}(0) \,
\hat{\cal M}_J\,
\delta^2(k_J^\perp - \hat{k}_J^\perp)\delta(\w_J-\bnP_J){\cal B}_{n_J}^{\mu',c_1'}(0) |0 \rangle \,.
\eea
To reach the formula above, we have summed over the final states using
$\sum_X |X\rangle \langle X| = 1$ and have re-inserted the measurement operators $\hat{\cal M}_A$.  We next define the beam and the jet functions
\bea\label{functions}
&&f_{\perp g/p}(z,p_T^{veto},R) = \, 
\int\mathrm{d}^2k^\perp\sum_{spin} -\w \langle p|{\cal B}_{n}^{\alpha,a}(0)\,
\hat{\cal M}_B\,
\delta^2(k^\perp-\hat{k}^\perp)\delta(\w-\bnP){\cal B}_{n,\alpha,a}(0) |p\rangle \,,\nn\\
&&J(R)g_\perp^{\mu\mu'}\delta^{cc'}= \,
 2 (2\pi)^3 (-\w_J)
\langle 0|{\cal B}_{n_J}^{\mu,c}(0)\,
\hat{\cal M}_J
\delta^2(k_J^\perp - \hat{k}_J^\perp)\delta(\w_J-\bnP_J){\cal B}_{n_J}^{\mu',c'}(0) |0 \rangle \,, 
\eea
and the soft function
\bea
S({\bf n}_J,R,p_T^{veto}) = \int\mathrm{d}^4k_s  \langle 0| S^{aa_1}_{n_a}S^{bb_1}_{n_b}S^{cc_1}_{n_J}(0)\,
\hat{\cal M}_s\,
\delta^4(k_s-\hat{k}_s) S^{a_1a'}_{n_a}S^{b_1b'}_{n_b}S^{c_1c'}_{n_J}(0)|0\rangle \,.
\eea
We emphasize that the beam and jet functions are well-defined only after
the soft zero-bin subtraction has been properly performed~\cite{Manohar:2006nz}.

Finally, we arrive at our result
\bea\label{facthiggs}
\mathrm{d}\sigma &=& \mathrm{d}\Phi_H \mathrm{d}\Phi_J\,\int\mathrm{d}x_a \mathrm{d} x_b\,
\frac{1}{2\hat{s}} \frac{1}{4}\left(\frac{1}{N_c^2-1}\right)^2\,
 (2\pi)^4 \delta^4\left(q_a + q_b - q_J -q_H\right)\nn\\
&&\hspace{5.ex}\times {\rm Tr}(H \cdot S )\,
f_{\perp g/p_a}(x_a,p_T^{veto},R) 
f_{\perp g/p_b}(x_b,p_T^{veto},R)
J(R) \,,
\eea
with the help of identifying
$\sum_n \frac{1}{2(2\pi)^3\w_J}\mathrm{d}\w_J \mathrm{d}^2k_J^\perp$
as the massless particle phase space
$\mathrm{d}\Phi_J = \frac{{\bar q}_J}{8(2\pi)^3}\mathrm{d}q_J\mathrm{d}\Omega$.
We define the hard function $H \equiv C C^\dagger$.  The trace
is over the color indices. Since $p_T^{veto}$ is much larger
than $\Lambda_{QCD}$, the beam function can be further matched onto
parton distribution functions at the beam scale $\mu_B \sim p_T^{veto}$:
$f_{i,\perp}  = {\cal I}_{ij} \otimes f_j(x)$.  The matching 
coefficient ${\cal I}_{ij}$ can be calculated perturbatively. 
We have chosen our normalization so that 
at leading order, ${\cal I}_{ij} = \delta(1-x)\delta_{ij}$, $J(R) = 1$ and
$S = \delta^{aa'}\delta^{bb'}\delta^{cc'}$.  The factorization reproduces the tree-level $gg\to Hg$ cross section. We note
that for NLL resummation, only the leading-order matching coefficients are needed. In the Appendix, we will present the NLO results for the jet and beam functions, up to ${\cal O}(R)$ corrections.

The formalism is easily generalized to processes with an arbitrary number
of jets and non-strongly interacting particles in the final state.  All
the arguments go through identically.  We find
\bea\label{factgen}
\mathrm{d}\sigma &=& \mathrm{d}\Phi_{H_c}\mathrm{d}\Phi_{J_i}\,
{\cal F}(\Phi_{H_c},\Phi_{J_i})
\,
\sum_{a,b}\int \mathrm{d}x_{a} \mathrm{d}x_b \frac{1}{2\hat{s}}\,
 (2\pi)^4 \delta^4\left(q_a + q_b - \sum_i^nq_{J_i} -\sum_cq_{H_c}\right)\nn\\
&&\times 
\bar{\sum_{\rm spin}}
\bar{\sum_{\rm color}}
{\rm Tr}(H\cdot S)\,
{\cal I}_{a,i_aj_a} \otimes f_{j_a}(x_a)\,
{\cal I}_{b,i_bj_b} \otimes f_{j_b}(x_b)
\prod_i^n J_{J_i}(R)\,,
\eea
where $\mathrm{d}\Phi_{Hc}$ and $\mathrm{d}\Phi_{j_i}$
denote the phase space measures for the color neutral particle $H_c$ and 
the massless jets $J_i$, respectively. ${\cal F}(\Phi_{H_c},\Phi_{J_i})$ includes all additional
phase-space cuts other than the $p_T$ veto acting on $H_c$ and the $n$ hard jets, which should
guarantee well separated $n$-jet final states ($n_{J_i}\cdot n_{J_j} \gg \lambda$).  The measured jet $p_T^J$ should be much larger than $p_T^{veto}$. 

We note that possible issues arise when attempting to extend the resummation to the NNLL level, as pointed out in Ref.~\cite{Tackmann:2012bt}.  For $R \gg \lambda$, corrections of the form
\begin{equation}
\alpha_s^2 \,R^2 \,\text{ln} \,\lambda
\label{prob1}
\end{equation}
appear, which prohibit soft-collinear factorization.  In the limit $R \sim \lambda$, clustering logarithms of the form
\begin{equation}
\alpha_s^2 \,\text{ln} \, R \,\, \text{ln} \,\lambda
\label{prob2}
\end{equation}
which prohibit even NLL resummation occur.  Let us study the numerical impact of these terms for the parameter values relevant for Higgs production.  Setting $R=0.4$, $m_H=126$ GeV, and $p_T^{veto}=25$ GeV, we find
$R^2 = 0.16$, $\text{ln} \, (1/\lambda) =1.6$, and $\text{ln} \, (1/R) = 0.92$.  For $R=0.5$,  $\text{ln} \, (1/R) = 0.69$.  It is clear that the corrections of Eq.~(\ref{prob1}) are indeed power-suppressed for experimentally-relevant value of $R\approx 0.4-0.5$.  They can be obtained to sufficient accuracy by matching to fixed-order results.  There is also a hierarchy $\text{ln} \, (1/\lambda) > \text{ln} \, (1/R)$; the clustering logarithms are not large for relevant jet parameters.  An eventual inclusion of the leading clustering effects by combining the resummation with a NNLO calculation of Higgs plus one-jet production should be sufficient. \OMIT{ It may also be possible to improve the accuracy of the $\text{ln}\, R$ terms using a refactorization ansatz for the soft function~\cite{Ellis:2009wj,Chien:2012ur}.}  We therefore believe that it makes sense to study the resummation of jet-veto logarithms in Higgs production, and proceed with our analysis.

\subsection{NLL Resummation}

Each ingredient in Eqs.~(\ref{facthiggs}) and~(\ref{factgen}) describes
fluctuations with a particular momentum scaling.  When the hard, jet,
beam and soft functions are calculated near their natural scales, no large
logarithms will arise. The typical scales for each sector
are
\bea
\mu_H \sim p_T^J\,,
\hspace{3.ex}
\mu_J \sim p_T^J R\,,
\hspace{3.ex}
\mu_B \sim \mu_S \sim p_T^{veto}\,.
\eea
However, calculating the cross section requires all the functions to
be evaluated at the same factorization scale $\mu$, which generates large
logarithms of the following ratios:
\bea
\frac{p_T^J}{\mu}\,,
\hspace{3.ex}
\frac{ p_T^J R}{\mu}\,,
\hspace{3.ex}
\frac{ p_T^{veto}}{\mu}\,.
\eea
These large logarithms can be resummed by evolving each function to the scale $\mu$ using the renormalization group  (RG) equation
\bea
\mu\frac{\mathrm{d}F}{\mathrm{d}\mu} = 
\gamma_\mu(\mu) F(\mu)\,.
\eea
The anomalous dimension $\gamma_\mu$ can be extracted most easily from the
$\epsilon$ poles of each function calculated using dimensional regularization.

However, like for small-$q_T$ resummation, a subtlety arises because of the identical virtuality shared by the collinear and soft degrees of freedom~\cite{Manohar:2006nz, Chiu:2012ir}. Various efforts have been made in SCET to regulate
this rapidity divergence~\cite{Chiu:2012ir, Mantry:2009qz, Becher:2011dz}, which all have shown be able
to reproduce correctly the NLO fixed-order QCD singularities. In our current
approach, we adopt the formalism proposed in Ref.~\cite{Chiu:2012ir}.  We regulate the extra divergence by modifying the collinear and the soft Wilson lines as follows:
\bea
&&W_n \to \sum_{\rm perm.}\,
\exp\left(-g_s\frac{1}{\bar {\cal P}}\,
\left[w^2 \frac{|{\bar {\cal P}}|^{-\eta}}{\nu^{-\eta}}
\bn \cdot A_n\right] \right)\,, \nn \\
&&S_n \to  \sum_{\rm perm.}\,
\exp\left(-g_s\frac{1}{{\cal P}}\,
\left[w \frac{|2{ {\cal P}}^3|^{-\eta/2}}{\nu^{-\eta/2}}
n \cdot A_s\right] \right) \,,
\eea
where the bookkeeping parameter $w$ will be set to $1$ as $\eta \to 0$. 
The effective rapidity cut-off $\nu$ and the new parameter
$\eta$ play similar roles as $\mu$ and $\epsilon$ in dimensional
regularization.  The corresponding rapidity-RG equation commutes
with the conventional one for the beam and the soft functions:
$[\mu\frac{\partial}{\partial\mu}, \nu\frac{\partial}{\partial\nu}] = 0$,  
\bea
\nu\frac{\mathrm{d}F_{B,S}}{\mathrm{d}\nu} = 
\gamma_\nu(\nu) F_{B,S}(\nu)\,.
\eea
The natural $\nu$-scale choices for the beam and the soft functions are
\bea
\nu_B \sim x_{a,b}\sqrt{s}\,,
\hspace{3.ex}
\nu_S \sim p_T^{veto} \,.
\eea
Since the physical cross section is $\mu$ and $\nu$ independent, 
the anomalous dimensions must obey the consistency conditions
\bea
&&\gamma_H^\mu + \gamma_J^\mu + \gamma_B^\mu + \gamma_S^\mu = 0 \,, \nn \\
&&\gamma_B^\nu + \gamma_S^\nu = 0 \,,
\label{consistency}
\eea
up to power corrections of order 
$\lambda$ and effects due to finite jet (beam) separations~\cite{Ellis:2009wj}. We will use these conditions to extract the anomalous dimension for
the soft function.  The general solution to the RG equation can be formally written as
\bea
F(\mu,\nu) = U(\mu,\nu,\mu_0,\nu_0) F(\mu_0,\nu_0) \,.
\eea
The explicit form of the evolution kernel $U$ for each function along with all details needed for NLL resummation are given in the Appendix.

Recalling that for NLL resummation only the tree-level Wilson
coefficients are needed, we find the following simplified expression for production of a color-neutral particle plus one jet:
\bea\label{resumhiggs}
\mathrm{d}\sigma_{\rm NLL}
 &=& \sum_{ab}\int \mathrm{d}\hat{\sigma}^{ab\to h k}_{\rm LO}(\mu_H,x_a,x_b)\,
f_a(\mu_B,x_a)f_b(\mu_B,x_b) \nn \\
&& \times U_{H,k}(\mu,\mu_H)\,
U_{S,k}(\mu,\nu,\mu_S,\nu_S)\,
{\cal I}_{B,a,b}(\mu,\nu,\mu_B,\nu_B,x_a,x_b)\,
{\cal R}_{J}(\mu,\mu_J,R)\,.
\eea
For a more general process, the evolutions of the hard and 
the soft functions will usually induce operator mixing in color space. Therefore, the NLL
cross section for multi-jet production reads
\bea\label{resumgen}
\mathrm{d}\sigma_{\rm NLL}
 &=& \sum_{ab}\int \mathrm{d}x_a \mathrm{d}x_b\,
{\rm Tr}\left[H^{ab\to h_c \{k\}}_{\rm LO}(\mu_H)\,S\,
U_{H,\{k\}}(\mu,\mu_H)\,
U_{S,\{k\}}(\mu,\nu,\mu_S,\nu_S)\right]
 \nn \\
&& \times \,
f_a(\mu_B,x_a)f_b(\mu_B,x_b)
{\cal I}_{B,a,b}(\mu,\nu,\mu_B,\nu_B,x_a,x_b)\,
\prod_{i}{\cal R}_{J_i}(\mu,\mu_J,R)\,.
\eea

We note that due to the separation between the 
scales inside and outside the allowed jets, a full refactorization or a 
refactorization ansatz of the soft
function~\cite{Ellis:2009wj,Chien:2012ur} may be helpful in improving the resummation
accuracy.  Since this involves $\text{ln} \, R$ effects and is therefore moderate for the experimentally-interesting $p_T^{veto}$ and $R$, its will
be left for further studies.  There also exist
non-global logarithms~\cite{Dasgupta:2001sh} beginning at the NNLL level whose
resummation is beyond the scope of our formalism presented in this work.

Different schemes are used in the literature to determine the accuracy of resummation prescriptions.  When calling our result NLL, we use the order-counting defined in Ref.~\cite{Berger:2010xi}.  Denoting the large logarithms associated with the veto scale generically as $L$, our NLL captures the two leading logarithms at each order in $\alpha_s$.   We correctly obtain 
$\alpha_s L^2$ and $\alpha_s L$, $\alpha_s^2 L^4$ and $\alpha_s^2 L^3$, $\alpha_s^3 L^6$ and $\alpha_s^3 L^5$, and so on.  To obtain the next tower of logarithms ($\alpha_s^2 L^2$, $\alpha_s^3 L^4$, etc.), we would need to include the jet, beam and soft functions at 1-loop.  This would correspond to NLL' in the language of Ref.~\cite{Berger:2010xi}.  All ingredients are currently known for this extension except for the one-loop soft function, which is not difficult to obtain.  We plan to include this term in future detailed numerical studies.  Control over the next tower of logarithms ($\alpha_s^2 L$, $\alpha_s^3 L^3$, etc.) would require the two-loop non-cusp anomalous dimensions $\gamma_x$, and would correspond to NNLL in the notation of Ref.~\cite{Berger:2010xi}.

\section{Numerical results for Higgs+jet production}\label{sechiggs}

We present in this section numerical results for $pp\to h + \text{jet}$ at 
an $8 {\rm TeV}$ LHC with a jet veto imposed.  We combine our 
resummation with the NLO cross section from MCFM to produce NLL+NLO results.  In the numerics presented here we restrict the leading jet rapidity to $|\eta_J|<2.5$ and veto all other jets with $p_{T,i}>p_T^{veto}$ and $|\eta_i|<\infty $ for simplicity.  Experiments typically only veto jets in the range $|\eta| \lesssim 4.5$.  However, we expect that this boundary effect will be small~\cite{Banfi:2012yh}.  It is simple to include this constraint if desired, as discussed in the previous section. We have included the $gg$, $qg$, and $q\bar{q}$ partonic scattering channels in obtaining these results.  All relevant beam, jet and soft functions, as well as anomalous dimensions needed for this calculation, are presented in the Appendix.

\begin{figure}[!ht]
\begin{center}
  \includegraphics[width=4.5in,angle=0]{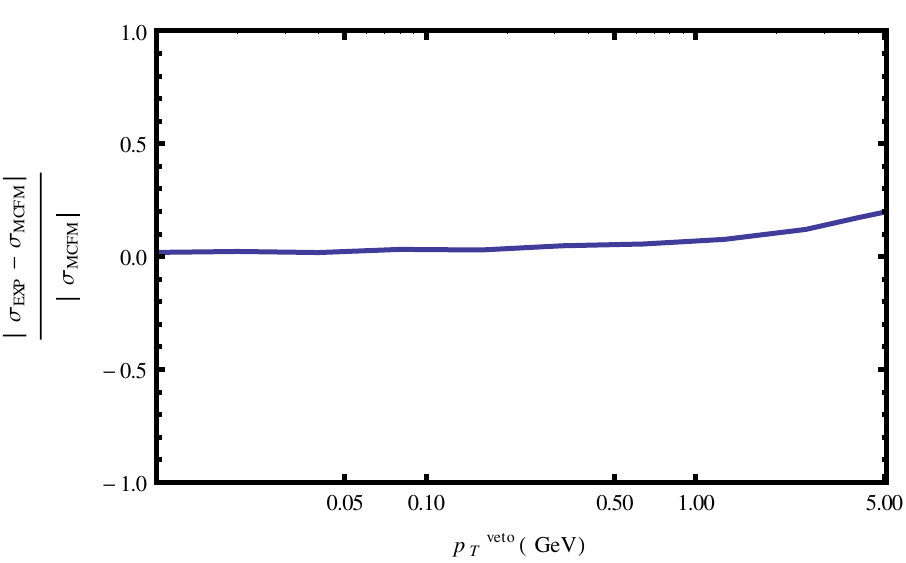}
\end{center}
\vspace{-0.5cm}
\caption{Presented is the ratio of the  expanded SCET cross section $|\sigma_{\rm exp}|$ and the full NLO QCD calculation 
from MCFM $\sigma_{\rm MCFM}$, $|\sigma_{\rm exp}-\sigma_{\rm MCFM}|/|\sigma_{\rm MCFM}|$.
We have required the leading jet 
$p^J_T > 120{\rm GeV}$, $|\eta_J| < 2.5$, have set $R = 0.1$ and have made
$p_T^{veto}$ as low as $0.01{\rm GeV}$. The excellent agreement
between the expanded SCET prediction and the MCFM cross section for such low
$p_T^{veto}$ implies that our formalism catches all the NLL structures. }\label{SCETMCFM}
\end{figure}

We begin by demonstrating that our formalism correctly sums all the next-to-leading logarithms
of $p_T^{veto}/p_T^J$ by comparing the expanded NLL production rate 
with the MCFM NLO result~\cite{Campbell:2010ff}. In the expanded NLL result, we have included the large non-logarithmic virtual corrections by using the full
NLO hard function taken from Ref.~\cite{Schmidt:1997wr}.
The validity of our
formalism is shown in Fig.~\ref{SCETMCFM}, where we show the agreement between these two results in the region where $\log p_T^{J}/p_T^{veto}$ becomes large.  This demonstration is based on the dominance of the 
log terms over the other contributions omitted in SCET in the small $p_T^{veto}$ region.

We next study the cross section for Higgs+jet production. Since Eq.~(\ref{resumhiggs}) is only valid for $p_T^{veto} \ll p_T^J$, in order to give a prediction over the entire allowed kinematic range, we have to combine
the resummed formula of Eq.~(\ref{resumhiggs}) with the full NLO result. For this purpose,
we adopt the matching scheme proposed in Ref.~\cite{Banfi:2012yh}, in which
the RG-improved cross section is taken as 
\bea\label{RGimprove}
\sigma = \,
\left( \frac{\sigma_{\rm NLL}}{\sigma_{\rm LO} } \right)^Z\,
\Big[ \sigma_{\rm NLO}(\mu) - Z \left(\sigma_{\rm exp}(\mu)-\sigma_{\rm LO}(\mu) \right) \Big]\,,
\label{fixedmatch}
\eea
where $Z = \left(1-p_T^{veto}/p_{T,veto}^{max} \right)$. $\sigma_{\rm LO}$ is the LO cross section and $\sigma_{\rm NLO}$ is the cross section calculated through NLO
using MCFM.
$\sigma_{\rm exp}$ is obtained by expanding the resummed cross section $\sigma_{\rm NLL}$ in Eq.~(\ref{resumhiggs}).  
We postpone rigorous study of the uncertainty induced by the choice of matching procedure to future work.  In evaluating the resummed production rate
$\sigma_{\rm NLL}$, we fix
\bea
&&\mu_H = \Big[(x_a\sqrt{s})^{T_a\cdot T_a}( x_b \sqrt{s})^{T_b\cdot T_b}\,
( p_T^J )^{T_J\cdot T_J}\Big]^{\frac{1}{\sum_iT_i\cdot T_i}}\,,\nn \\
&&\mu_J = p_T^J R \,, \nn \\
&&\mu_B = \mu_S = p_T^{veto} \,. 
\eea
These choices minimize the logarithmic dependence in the hard, jet and the beam functions.  When we use this set of scale choices, the cross section
is also independent of the rapidity scale $\nu$.
\begin{figure}[h!]
\begin{center}
\includegraphics[width=5.0in,angle=90]{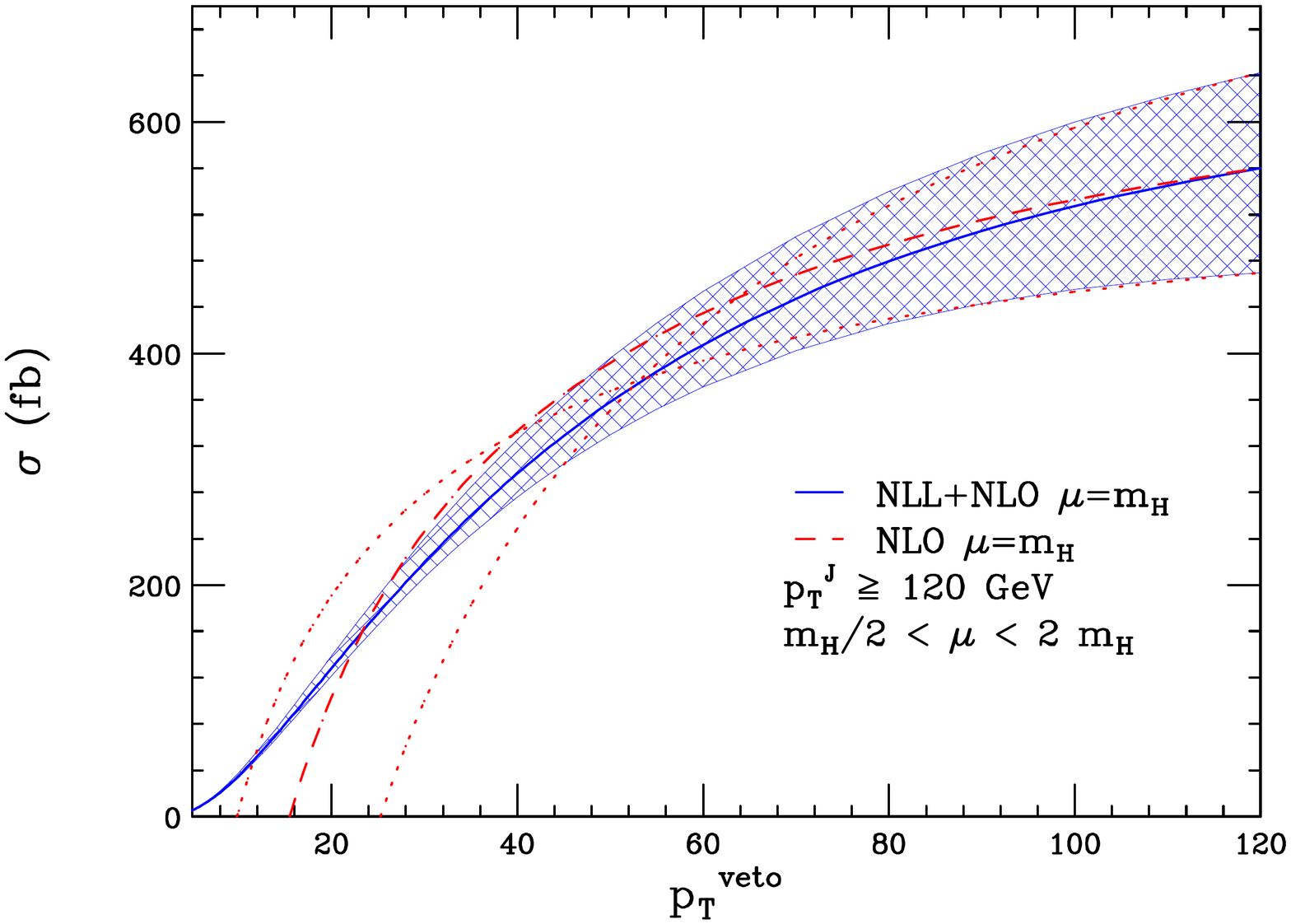}
\end{center}
\vspace{-1.0cm}
\caption{Shown are the NLO$+$NLL prediction and the NLO cross section as a function of $p_T^{veto}$ for $p_T^J \ge 120 $ GeV. The solid-blue
curve represents the RG-improved cross section in Eq.~(\ref{RGimprove}).  The 
dashed-red line shows the NLO result for the scale choice $\mu=m_H$, while the dotted red lines show the NLO result for $\mu=m_H/2$ and $\mu=2 m_H$ . The blue band reflects the scale uncertainty of the RG-improved rate. The band boundaries are set by the values at the scale choices $m_H/2$ and $2m_H$.}
\label{NLONLL}
\end{figure}

In Fig.~\ref{NLONLL}, we demonstrate the improvement obtained with NLL resummation
by showing the dependence of both the NLO and RG-improved integrated cross sections on different choices of $p_T^{veto}$.  We have set $p_T^J > 120{\rm GeV}$, $\mu= m_H = 126{\rm GeV}$ and have taken the anti-$k_T$ parameter
$R = 0.4$. For the matching in Eq.~(\ref{fixedmatch}), we use $p_{T,veto}^{max} = 120$ GeV in $Z$. We use the MSTW2008 NLO PDF set~\cite{Martin:2009bu} with two-loop $\alpha_s$ running.  We can see that for small values of $p_T^{veto}$,
the fixed-order cross section becomes negative while the ${\rm NLO}+ {\rm NLL}$ result remains positive. We also vary the scale $\mu$ from 
$m_H/2$ to $2m_H$ to estimate the theoretical uncertainty. We can see that resumming the large logarithms greatly reduces the residual scale variation for the experimentally relevant values $p_T^{veto} \approx 25-30$ GeV, leading to a more reliable prediction.  The fixed-order cross section exhibits little scale variation for $p_T^{veto} \approx 55$ GeV.  Similar behavior is observed for the Higgs plus zero-jet cross section~\cite{Anastasiou:2008ik,Stewart:2011cf}, and was argued to result from an accidental cancellation between several higher-order corrections with different origins.  The same argument holds here for the Higgs plus one-jet result.  As $p_T^{veto}$ becomes large, the fixed-order and resumed cross section coincide.  Since the separation between the hard scale and $p_T^{veto}$ vanishes in this limit, this behavior is expected.

\begin{figure}[h!]
\begin{center}
\includegraphics[width=5.0in,angle=90]{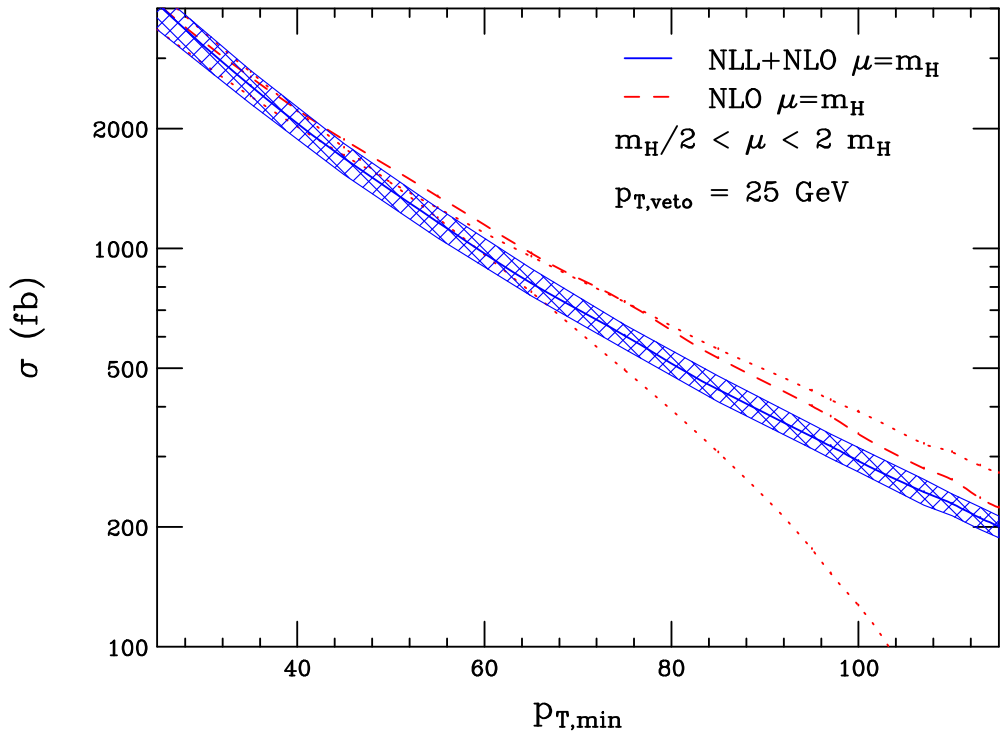}
\end{center}
\vspace{-1.0cm}
\caption{The NLO$+$NLL prediction and the NLO cross section as a function of $p_T^{min}$ for $p_T^{veto} =25 $GeV. The solid-blue
curve represents the RG-improved cross section in Eq.~(\ref{RGimprove}).  The 
dashed-red line shows the NLO result for the scale choice $\mu=m_H$, while the dotted red lines show the NLO result for $\mu=m_H/2$ and $\mu=2 m_H$ . The blue band reflects the scale uncertainty of the RG-improved rate. The band boundaries are set by the values at the scale choices $m_H/2$ and $2m_H$.}
\label{NLONLLpT}
\end{figure}

To gain some intuition regarding how low in $p_T^J$ the resummation of jet-veto logarithms leads to a difference from fixed-order, we fix $p_T^{veto} = 25$ GeV and integrate over the leading-jet transverse momentum subject to the constraint 
$p_T^J \geq p_{T,min}^j$.  We stress that some caution must be exercised in using these results.  Our effective theory framework is only valid when the hierarchy $p_T^{veto} \ll p_T^J$ exists.  When $p_T^{veto} \sim p_T^J$, our result reduces to the fixed-order result, which contains the ratio of scales $p_T^{veto} \ll m_H$.  A different effective theory framework consisting is needed to resume logarithms of this ratio.  With these caveats stated, we plot in Fig.~\ref{NLONLLpT} the NLO and NLL$+$NLO results as a function of the minimum allowed jet $p_T$ for $p_T^{veto} =25$ GeV.  Significant differences between both the central values and residual scale uncertainties persist down to low values of $p_T^{min}$, indicating the need to augment the fixed-order results with resummation over the entire $p_T^J$ region. \OMIT{ The vanishing of the scale variation band for both the NLL$+$NLO and NLO results for $p_T^{min} \sim 40$ GeV indicates the continued need for resummation in this region, and also the need to switch to a different effective-theory description for low $p_T^{min}$.}

\section{Conclusions}\label{seccon}

In this manuscript, we have established a formalism for the production
of color-singlet particles produced in association with an exclusive number of jets at the LHC. Using effective field theory techniques, we have proven the factorization theorem of Eqs.~(\ref{facthiggs}) and~(\ref{factgen}), which allows us
to resum large Sudakov logarithms of the form 
$\log p_T^{veto}/p_T^J$ to all orders.  We have focused on Higgs production in association with a jet as an example.  We have demonstrated
by the excellent agreement between the expanded NLL result and NLO QCD cross section that our formalism correctly captures all relevant large logarithms in the $p_T^{veto} \to 0$ limit.  The scale uncertainty of the cross section is greatly reduced by inclusion of the resummation.  By matching our results with MCFM, we provide a NLL+NLO result for Higgs+jet production valid over the entire kinematic range.  With our results, it is easy to supplement fixed-order results for vector boson or Higgs boson production with the resummation of jet-veto logarithms in order to provide predictions valid throughout phase space. 

Several future directions remain to be pursued. With our results it is possible to improve the theoretical predictions for Higgs plus one or two jets.  This will be of great phenomenological importance as the properties of the new state discovered at the LHC are further analyzed.  We plan to further investigate these phenomenological application of our formalism.  It is also interesting to study the effective theory valid when both the veto scale and the leading jet $p_T$ are smaller than $m_H$.  Beyond NLL, extra 
clustering effects will enter the cross section, which are 
not completely understood~\cite{Tackmann:2012bt} yet.   We have argued that the numerical impact of these terms should 
be subdominant to the effects studied here.  However, a NNLO calculation
of the jet, beam and soft functions would be helpful in determining whether
this formalism can be extended beyond NLL. In the current work,
we have simplified the calculation of the 
anti-$k_T$ jet function by keeping the leading $R$ contributions only in the resummation formulae. It will
be phenomenologically interesting to include higher order corrections in $R$
to improve the accuracy.  We have also neglected for simplicity the boundary effects due
to the experimentally finite $\eta$ range.  These issues will be addressed in future detailed phenomenological studies.

\section*{Acknowledgments}
We thank Kirill Melnikov, Jui-yu Chiu, Sonny Mantry, Markus Schulze, 
Jonathan Walsh, Frank Tackmann and Giulia Zanderighi for helpful discussions. We thank Joey Huston to point out a numerical bug in 
Fig.~\ref{NLONLLpT} in the previous verison of this work.
This work is supported by the U.S. Department of Energy, Division of High Energy Physics, under contract DE-AC02-06CH11357 and the grants DE-FG02-95ER40896 and DE-FG02-08ER4153.

\appendix
\section{Fixed-order jet and beam functions}
In this Appendix, we list all ingredients needed for NLL resummation. We start
with the NLO calculation of the jet and the beam functions, whose definitions
can be found in Eq.~(\ref{functions}).  The anti-$k_T$ jet function is calculated using the measurement function
\bea
\hat{\cal M}_J = \Theta(\Delta \eta_{ij}^2 + \Delta \phi_{ij}^2 < R^2 )
+{\cal O}(p_T^{veto})\,.
\eea
We explicitly calculate that the jet collinear radiation
leaking outside the jet is power-suppressed by $p_T^{veto}$ after correctly subtracting the soft zero-bin contributions.  Since numerically $R\ll 1$, we can simplify the measure using
\bea
\Delta\eta_{ij}^2 + \Delta \phi_{ij}^2 = 
2\cosh(\Delta\eta_{ij}  ) - 2\cos(\Delta \phi_{ij}) + {\cal O}(R^4).
\eea
In this limit, the NLO jet functions for gluons and quarks become
\bea
J^{(1)}_g &=& \,
\frac{\alpha_s(\mu)}{2\pi}\left[C_A\left(\frac{67}{9}-\frac{3\pi^2}{4} \right) \,
-\frac{23}{9}\frac{n_f}{2}\,
+\beta_0\log \frac{\mu}{p_T^JR}\,
+2C_A \log^2 \frac{\mu}{p_T^J R} \,
\right] +{\cal O}(R^2) \,, \nn \\
J^{(1)}_q &=& \,
\frac{\alpha_s(\mu)}{2\pi}C_F\,
\left[\frac{13}{2}-\frac{3\pi^2}{4}
+3\log \frac{\mu}{p_T^JR}\,
+2\log^2\frac{\mu}{p_T^J R} \,
\right] +{\cal O}(R^2) \,.
\eea

The measure for the beam function with one emission is 
\bea
\hat{\cal M}_B = \Theta\left(k_{T,i} < p_T^{veto}\right)\, 
\Theta\left(|\eta_i| < \eta_{\rm cut} \right) \,
+ \Theta(|\eta_i| > \eta_{\rm cut} )\,.
\eea
Experimentally, $\eta_{\rm cut} \sim 4.5$.  For simplicity, we set
$\eta_{\rm cut} = \infty$ here.  We note that this difference does not affect the anomalous dimension of the beam function, it only changes the finite part. 
The calculation is performed using the 't Hooft-Veltmann scheme.
The NLO matching coefficient ${\cal I}$ for the beam function is found to be
\bea
{\cal I}^{(1)}_{gg}(z) &=& \,
\frac{\alpha_s(\mu)C_A}{2\pi}\left(\,
4\log\frac{\mu}{p_T^{veto}}\log\frac{\nu}{\bnp} \delta(1-z)
-2\tilde{p}_{gg}(z) \log \frac{\mu}{p_T^{veto}} 
 \right)\,, \nn \\
{\cal I}^{(1)}_{qq}(z) &=& \,
\frac{\alpha_s(\mu)C_F}{2\pi}\left(\,
4\log\frac{\mu}{p_T^{veto}}\log\frac{\nu}{\bnp}\delta(1-z)\,
-2\tilde{p}_{qq}(z) \log \frac{\mu}{p_T^{veto}}\,
+(1-z)
\right)\,, \nn \\
{\cal I}^{(1)}_{gq}(z) &=& \frac{\alpha_s(\mu)C_F}{2\pi}\left(\,
-2p_{gq}(z)\log\frac{\mu}{p_T^{veto}} + z 
\right)\,, \nn \\
{\cal I}^{(1)}_{qg}(z) &=& \frac{\alpha_s(\mu)T_F}{2\pi}\left(\,
-2p_{qg}(z)\log\frac{\mu}{p_T^{veto}} + 2z(1-z) 
\right) \,,
\eea
with 
\bea
&&\tilde{p}_{gg}(z) = \frac{2z}{(1-z)_+}+2z(1-z)+2\frac{1-z}{z}\,, \nn\\
&&\tilde{p}_{qq}(z) = \frac{1+z^2}{(1-z)_+}  \,, \nn \\
&&p_{gq}(z) = \frac{1+(1-z)^2}{z} \,, \nn \\
&&p_{qg}(z) = 1 -2z +2z^2\,.
\eea
From these fixed order calculations, we can determine the anomalous dimensions
used for RG evolution. The anomalous dimension for the jet function is given by
\bea
\gamma_{J_i} = 2\Gamma_{\rm cusp} 
T_i^2 \log \frac{\mu}{p_T^{J_i} R} +\gamma_{J_i}\,.
\eea
For the beam function, it is
\bea
\gamma_B^\nu &=& 2\Gamma_{\rm cusp} T_i^2 \log \frac{\mu}{p_T^{veto}} \,, \nn\\
\gamma_B^\mu &=& 2\Gamma_{\rm cusp} T_i^2 \log \frac{\nu}{\bnp} + \gamma_{B_i}\,,
\eea
where $T_i^2 = C_A$ for gluon and $T_i^2 = C_F $ for quark.  Here, 
\begin{equation}
\Gamma_{\rm cusp} = \frac{\alpha_s}{4\pi}\Gamma_0 
+ \left(\frac{\alpha_s}{4\pi}\right)^2\Gamma_1 + \dots .
\end{equation}

\section{RG evolution}
The evolutions of the jet and the beam functions are given by
\bea
{\cal R}_{J_i} &=& \exp\left[-2T_i^2 S(\mu_J,\mu)-A_{J_i}(\mu_J,\mu) \right]
\left( \frac{\mu_J}{p_T^{J_i}R}\right)^{-2T_J^2 A_\Gamma(\mu_J,\mu)} \,, \nn \\
U_{B,a} &=& 
\exp\left[-T_a^2 A_\Gamma(p_T^{veto},\mu)\log\frac{\nu^2}{\nu_B^2}\right]
\exp\left[-T_a^2 A_\Gamma(\mu_B,\mu)\log\frac{\nu^2_B}{\w_a^2}-A_{B_a}(\mu_B,\mu) \right]\,. \nn \\
\eea
We have defined ${\cal I}_{B,a,b} = U_{B,a}U_{B,b}$ in Eqs.~(\ref{resumhiggs}) and~(\ref{resumgen}). For the Higgs production process considered in this 
work, we have, the following color identities: for the $ggg$ channel, $T_i\cdot T_j = -C_A/2$; for the 
$q_1{\bar q}_2g_3$ channel, 
$T_1\cdot T_2 = -(C_F-C_A/2)$ and $T_1\cdot T_3 = T_2\cdot T_3 = -C_A/2$.  The anomalous dimension for the hard function can be found
in Ref.~\cite{Kelley:2010fn}.  The solution to the RG equation
is
\bea
U_H&=&\exp\left[2\sum_i T_i^2 S(\mu_H,\mu)-2A_H(\mu_H,\mu)\,
+2A_\Gamma(\mu_H,\mu) \sum_{i\neq j}\frac{T_i\cdot T_j}{2}\log\Delta R^2_{ij}\right]\nn\\
&& \times
 \prod_i \left(\frac{\mu_H}{\w_i} \right)^{2T_i^2A_\Gamma(\mu_H,\mu)}\,, 
\eea
where $\Delta R^2_{ij} = 2\left(\cosh(\Delta \eta_{ij})-\cos(\Delta\phi_{ij} \right))$ for $i,j \ne a,b$, $\Delta R^2_{ia} = e^{-\eta_i}$, $\Delta R^2_{ib} = e^{\eta_i}$ and $\Delta R^2_{ab} = 1$. Also $\w_i = p_T^{j_i}$ if $i\in J$, otherwise $\w_a = x_a\sqrt{s}$.
The soft anomalous dimension is determined by the consistency relation in Eq.~(\ref{consistency}),
which gives the solution
\bea
U_S& = &\exp\left[-2\sum_{i\in B}T_i^2 S(\mu_s,\mu) -A_s(\mu_s,\mu)\, 
- 2A_\Gamma(\mu_s,\mu)\sum_{i \ne j}\frac{T_i\cdot T_j}{2}\log \Delta R^2_{ij}\right]\nn\\
&& \times \left(\frac{1}{R}\right)^{\sum_{i\in J}2T_i^2A_{\Gamma}(\mu_s,\mu)}
\left(\frac{\nu_s}{\mu_s} \right)^{\sum_{i\in B}2T_i^2A_{\Gamma}(\mu_s,\mu)}
\left(\frac{\nu}{\nu_s} \right)^{\sum_{i\in B}2T_i^2A_{\Gamma}(p_T^{veto},\mu)}\,.
\eea

For the NLL resummation, we need 
\bea
A_{\Gamma}(\mu_i,\mu_f)&=& \frac{\Gamma_0}{2\beta_0}\left\{
\log r + \,
\frac{\alpha_s(\mu_i)}{4\pi}\left(\frac{\Gamma_1}{\Gamma_0}-\frac{\beta_1}{\beta_0} \right)\,
(r-1) \,
\right\}\,,
\eea
and 
\bea
S(\mu_i,\mu_f) &=& \frac{\Gamma_0}{4\beta_0^2}\left\{ 
\frac{4\pi}{\alpha_s(\mu_i)}\left(1-\frac{1}{r}-\log r\right)\,
+\left(\frac{\Gamma_1}{\Gamma_0}-\frac{\beta_1}{\beta_0} \right)(1-r+\log r)\,
+\frac{\beta_1}{2\beta_0}\log^2 r 
\right\}\,, \nn \\
\eea
where $r=\alpha_s(\mu_f)/\alpha_s(\mu_i)$ and 
\bea
&&\beta_0 = \frac{11}{3} C_A - \frac{4}{3} T_F n_f  \,,\\
&&\beta_1 = \frac{34}{3} C_A^2 - \frac{20}{3} C_A T_F n_f - 4 C_F T_F n_f \, , \\
&&\Gamma_0 = 4 \,,\\ 
&&\Gamma_1 = 4 \left[ C_A \left( \frac{67}{9} - \frac{\pi^2}{3} \right) -
\frac{20}{9} T_F n_f \right]\,.
\eea
$A_{J/B}$, $A_H$ and $A_S$ are needed at leading order, and can be obtained
by substituting the $\Gamma_0$ in $A_\Gamma$ with the corresponding
$\gamma_0$ and expanding in $\alpha_s$. For $A_{J/B}$ we have
\bea
&&\gamma_0^{J_q} =  6 C_F  \, , 
\eea
and
\bea
&&\gamma_0^{J_g} =  2\beta_0 \, ,
\eea
for quark and gluon jets or beam functions, respectively.  For $\gamma_H = \sum_i \gamma_i$, we have
\bea
   \gamma_0^q &=& -3 C_F \,, 
\eea
and
\bea
   \gamma_0^g &=& - \beta_0 \,,
\eea
The two loop $\gamma_S$ can be extracted via the relation 
$2\gamma_H + \gamma_{J_i} +\gamma_S+\gamma_{B_a}+\gamma_{B_b} = 0$. At one loop,
$\gamma_0^S = 0$.

\section{Expanded results}
Expanding out the resummed result will give us the fixed order singular term up to NLL.  We find
\bea
\frac{H}{H_0} = 1 +  \frac{\alpha_s(\mu)}{4\pi}\left(-\frac{\Gamma_0}{2}\sum_i T_i^2 \frac{L^2}{2}\,
+\left(-\Gamma_0 \sum_{i\ne j}\frac{T_i\cdot T_j}{2} \log\Delta R^2_{ij}\,
-\Gamma_0 \sum_i T_i^2 \log\frac{\mu_H}{\w_i} +\gamma_0^H +n\beta_0 
\right)L \right)\,, \nn \\
\eea
where we have assumed at the leading order $H_0 \propto \alpha_s^n$ and defined $L=\log(\mu/\mu_H)^2$ where
$\mu_H$ is of order $p_T^J$, and $\Delta R_{ij}^2$ and $\w_i$ have been defined previously. 

The expanded jet evolution gives
\bea
J_i = 1+\frac{\alpha_s(\mu)}{4\pi}\left(
\frac{\Gamma_0}{2}T_i^2 \frac{L^2}{2}
+\left(\Gamma_0T_i^2 \log \frac{\mu_J}{p_T^{J_i}R} +\frac{\gamma^{J_i}_0}{2} \right)L
\right)\,,
\eea
where $L = \log(\mu/\mu_j)^2$.  We have checked that it reproduces singular pieces of the fixed-order NLO calculation of the jet function. To see this we note that
the first two NLO terms can be combined to give
$\frac{\Gamma_0}{4}T_i^2 \log^2 \frac{\mu^2}{(p_T^{J_i}R)^2} + {\cal O}(\lambda^0) $ as long as $\mu_J$ is of order $p_T^{J_i} R$.

The singular terms of the beam coefficient are given by
\bea
{\cal I}_{aa'}(z) &=& \delta(1-z)\delta_{aa'}\left(1+\frac{\alpha_s(\mu)}{4\pi}
\left[ \Gamma_0 T_a^2 \left(
\log\frac{\mu}{p_T^{veto}} \log\frac{\nu^2}{\nu_B^2} 
+\log\frac{\mu}{\mu_B} \log\frac{\nu_B^2}{\w_a^2}
\right)+\gamma_0^{B_a} \log\frac{\mu}{\mu_B}
\right]
\right)\nn \\
&& - \frac{\alpha_s(\mu)}{\pi}p_{aa'}(z)\log\frac{\mu}{\mu_B} \,,
\eea
where $p_{aa'}(z)$ is the normal splitting function. 
This result again agrees with the fixed order calculation if we choose
$\mu_B \sim p_T^{veto}$ and $\nu_B \sim \w_a$.

Finally, the soft function is
\bea
S& = &1+\frac{\alpha_s(\mu)}{4\pi}\left(\,
\frac{\Gamma_0}{2}\sum_{a\in B}T_a^2
\left[
\frac{L^2}{2} 
+\log\frac{\nu_S^2}{\nu^2}\log\frac{\mu^2}{(p_T^{veto})^2}
\right]
\right)\nn \\
&& +\frac{\alpha_s(\mu)}{4\pi}\left(
\frac{\Gamma_0}{2}\sum_{i\in J}T_i^2\log R^2
+\frac{\Gamma_0}{2}\sum_{a\in B}T_a^2 \log\frac{\mu^2_S}{\nu^2_S}
+\Gamma_0\sum_{i\ne j}\frac{T_i\cdot T_j}{2}\log \Delta R_{ij}^2 + \frac{\gamma_0^S}{2}
\right)L \,, \nn \\
\eea
where $L=\log(\mu/\mu_S)^2$ and $\mu_S \sim\nu_S\sim p_T^{veto}$.
We note that once we combine the soft and the beam functions, the $\nu$ dependence in these two contributions cancels, as required. If we
set $\mu_B = \mu_S = p_T^{veto}$, the $\nu_S$ and $\nu_B$ dependence
also disappear.  We note that numerically, the $\log R^2 \times L$
term is of order $0.1 \times L$, which we interpret as a single-log term.

\end{document}